\documentclass[conference]{IEEEtran}
\IEEEoverridecommandlockouts
\usepackage{cite}
\usepackage{amsmath,amssymb,amsfonts}
\usepackage{algorithmic}
\usepackage{graphicx}
\usepackage{textcomp}
\usepackage{xcolor}
\usepackage{booktabs}
\usepackage{authblk} 

\def\BibTeX{{\rm B\kern-.05em{\sc i\kern-.025em b}\kern-.08em
    T\kern-.1667em\lower.7ex\hbox{E}\kern-.125emX}}
\begin{document}

\title{MamT$^4$: Multi-view Attention Networks for Mammography Cancer Classification\\}

\author[1]{Alisher Ibragimov}
\author[1]{Sofya Senotrusova}
\author[1]{Arsenii Litvinov}
\author[1]{Egor Ushakov}
\author[1]{Evgeny Karpulevich}
\author[1]{Yury Markin}

\affil[1]{Information Systems Department, ISP RAS, Russia}

\affil[ ]{\texttt{\{ibragimov,senotrusova,filashkov,ushakov,karpulevich,ustas\}@ispras.ru}}

\maketitle

\begin{abstract}
In this study, we introduce a novel method, called MamT$^4$, which is used for simultaneous analysis of four mammography images. 
A decision is made based on one image of a breast, with attention also devoted to three additional images: another view of the same breast and two images of the other breast. This approach enables the algorithm to closely replicate the practice of a radiologist who reviews the entire set of mammograms for a patient. Furthermore, this paper emphasizes the preprocessing of images, specifically proposing a cropping model (U-Net based on ResNet-34) to help the method remove image artifacts and focus on the breast region. To the best of our knowledge, this study is the first to achieve a ROC-AUC of 84.0 ± 1.7 and an F1 score of 56.0 ± 1.3 on an independent test dataset of Vietnam digital mammography (VinDr-Mammo), which is preprocessed with the cropping model.
\end{abstract}


\section{Introduction}

Breast cancer is a leading cause of cancer-related deaths among women~\cite{hopewood2018quality}. Regular screening is essential for early detection, with mammography being the primary screening tool~\cite{acr}. Mammography utilizes low-dose X-rays to detect tissue changes in the breast, making it effective in detecting malignancies like microcalcifications and clusters of calcifications~\cite{mammography}. Radiologists interpret mammograms based on standard terminology and the BI-RADS classification system, facilitating standardized reporting and risk assessment~\cite{birads}.

Although mammography is effective, it can result in false positives or negatives, requiring additional tests like biopsies~\cite{effectiveness}. To improve the efficiency of early screening, automated approaches in mammography, such as computer-aided diagnosis (CAD) systems, as well as solutions using machine learning and deep learning (DL) technologies, are being actively developed, assisting radiologists in interpreting mammogram~\cite{cad}.

Deep Learning has emerged as a highly effective method for image classification~\cite{LeCun}. Furthermore, DL has become one of the popular methods for detection of cancer pathologies, particularly, on mammograms~\cite{dl_mammography}.

A key aspect of mammographic examinations is the acquisition of images in different projections for each breast, requiring four images in total -- two for each breast (MLO and CC). Radiologists analyze the symmetry of lesions~\cite{symmetry}. This unique feature affects the training of DL models and the use of multi-view, a novel approach based on learning four projections at once, presented in this study.
Given the widespread significance of breast cancer diagnosis, this research on utilizing deep learning methods offers a valuable chance to enhance the automation of breast pathology detection and streamline the tasks of radiologists.

To sum up, the contributions of our paper are:
\begin{enumerate}
\item We propose MamT$^4$: a novel classification framework based on Transformer Encoder that utilizes feature vector representations from four views of mammography studies and outperforms single-view methods in classifying cancer status.

\item To the best of our knowledge, this paper is the first to achieve ROC-AUC of $84.0 \pm 1.7$ and F1 of $56.0 \pm 1.3$ on the VinDr-Mammo dataset (test subset).

\item As a preprocessing, we propose the following method: cropping the breast along its border using U-Net to increase the quality of classification.
\end{enumerate}

\section{Background and Preliminaries}
\label{sec:background}

\begin{figure}
  \centering
   \includegraphics[width=0.9\linewidth]{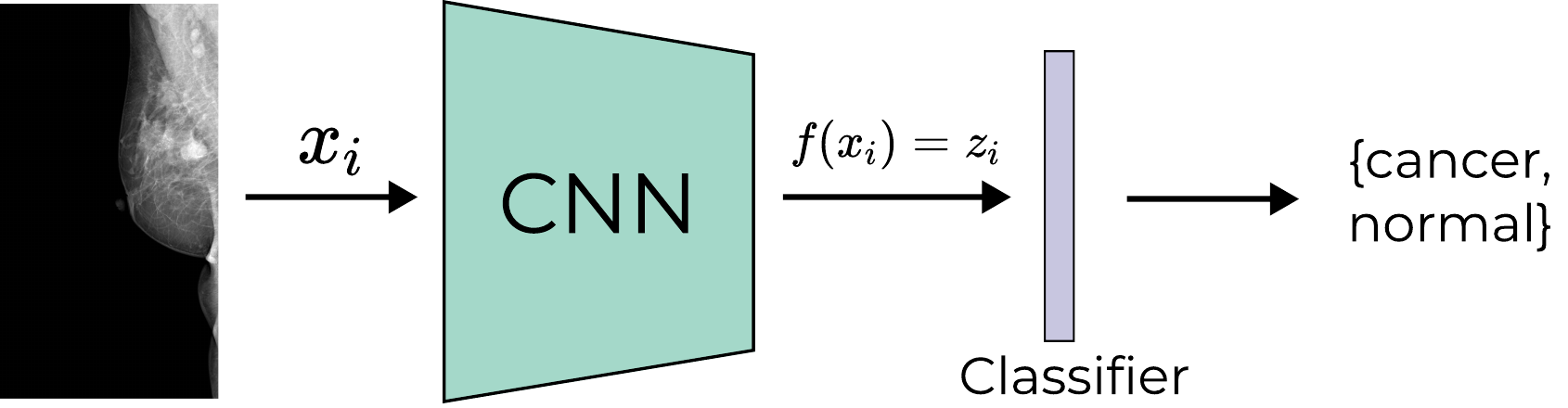}

   \caption{An overview of the training process of the CNN and classification layer applied to a binary classification problem using a single view. Subsequently, the trained CNN block is employed to derive a feature vector ($z_i$) from the mammography image ($x_i$).}
   \label{fig:single}
\end{figure}

\subsection{U-Net}
U-Net~\cite{DBLP:journals/corr/RonnebergerFB15}, introduced in 2015, is an encoder-decoder network tailored for semantic segmentation, which excels in medical image segmentation. The architecture efficiently maps low-resolution encoder features to high-resolution inputs through a decoder that uses pooling indices from the encoder for precise pixelwise classification. This setup enables U-Net to accurately delineate detailed features in medical images, crucial for identifying and segmenting various anatomical structures and abnormalities~\cite{ibragimov2023deep}. Its ability to handle small datasets effectively and its adaptability to various medical imaging modalities have made U-Net a standard choice in medical image analysis, enhancing diagnostic accuracy and aiding in clinical decision-making.

\subsection{Transformer Encoder}
\begin{figure*}[t]
  \centering
   \includegraphics[width=0.9\linewidth]{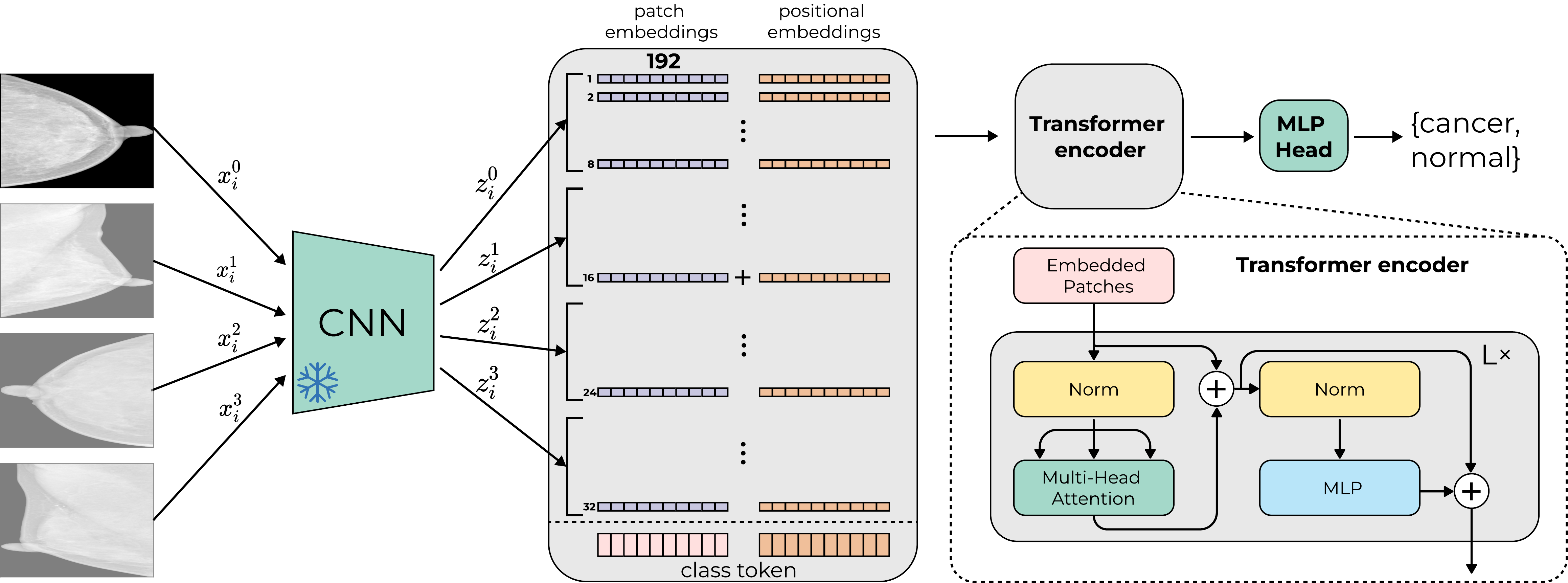}

   \caption{A summary of the MamT$^4$ framework that is used for cancer classification on $x_i^{0}$ images. Here, $x_i^{0}$ represents the primary view, while $x_i^{1}$ is the corresponding ipsilateral view to the primary one. Similarly, $x_i^{2}$ depicts the corresponding bilateral view to the main view, whereas $x_i^{3}$ illustrates the ipsilateral view of $x_i^{2}$. The CNN block, which gives the features vectors ($z_i^0$, $z_i^1$, $z_i^2$, $z_i^3$) with fixed length 1536 per each one, are untrainable during this stage.  
   Each vector is divided into fixed-size patches, each of which is linearly embedded. After adding position embeddings, the resulting sequence of vectors is fed to a Transformer Encoder.
   In order to perform classification, we use the standard approach of adding an extra learnable \texttt{[class]} token to the sequence. The illustration of the Transformer Encoder was inspired by Dosovitskiy et al.~\cite{dosovitskiy2020image}}
   \label{fig:snn}
\end{figure*}

Our approach draws inspiration from the ViT (Vision Transformer) framework~\cite{dosovitskiy2020image}. Transformer Encoder (TE) block consists of layer norm, multi-head self-attention (MSA) and multi-layer perceptron (MLP). Also, as shown in Figure~\ref{fig:snn} the TE block accepts combined embeddings as input. For all subsequent blocks, the inputs are the outputs from the previous block of the TE. 
There is a total of $L$ such TE blocks. 
Inside the TE, the inputs are first passed through a layer norm, and then fed to MSA layer with $N$ heads. Once we get the outputs from the MSA layer, these are added to the inputs (with skip connection) to get the outputs that again get passed to layer norm before being fed to the MLP block. The MLP consists of two linear layers and a GELU non-linearity. The outputs from the MLP block are again added to the inputs to get the final output from one TE block.

\subsection{Loss Function}
We use the~focal loss (FL) function was selected to achieve a greater stability when training on both frequent (normal cases) and rare (cancer cases) images~\cite{asgari2021deep}. 
For~the case of a binary classification, the focal loss can be written in the following form~\cite{lin2017focal}:

\begin{equation}
\mathrm{FL}\left(p_{\mathrm{t}}\right)=-\alpha_{\mathrm{t}}\left(1-p_{\mathrm{t}}\right)^\gamma \log \left(p_{\mathrm{t}}\right),
\end{equation}

\noindent where $\gamma \geq 0$ is a tunable focusing parameter, and $\alpha_t$ is a weighting factor for different classes to balance the importance for positive and negative examples. Namely, $\alpha_1=1-\frac{N_{c}}{N}$ for class \{cancer\} and $\alpha_0=1-\alpha_1$ for class \{normal\}, where $N_c$ is the number of images marked as \{cancer\} and $N$ is the total number of images in dataset.
For~notational convenience, $p_t$ was defined as:

\begin{equation}
p_{\mathrm{t}}= \begin{cases}p & \text { if } y=\mathrm{cancer} \\ 
1-p & \text { otherwise. }\end{cases}
\end{equation}

In the above, $y$ specifies the ground-truth class and $p \in [0, 1]$ is the model's estimated probability for the class with label $y = \mathrm{cancer}$.  This approach improved the performance of the small number of cancer images due to the modulating factor $(1 - p_t)^\gamma$.

\section{Methodology}
In this paper, we propose a novel method to achieve a better quality to classify mammography images.
To achieve such a goal, the two main components of our approach are presented in Section~\ref{sec:background}: image preprocessing using U-Net and Transformer Encoder to implement MamT$^4$. An overview of our proposed method is shown in Figure~\ref{fig:snn}.

\subsection{Crop mammogram}
The first step in image preprocessing is selecting the region of interest (ROI) through cropping. This process involves segmentation of large images to focus on specific areas that are of interest for further analysis, thereby making the task easier for subsequent neural network predictions. Cropping is useful as it helps concentrate the model's attention on relevant features without the distraction of background artifacts, which can be especially beneficial in datasets with limited examples~\cite{Abdelhafiz2019}. By selecting these regions and resizing them uniformly, models can more consistently learn from and recognize similar patterns in new data.

\subsection{Multi-view Analysis} 
\label{sect:mamt}

Inspired by previous works on an utilization of several images (Section~\ref{sec:multi_view}) to predict labels in classification tasks, we present a method that includes two training stages to construct a classification model, considering feature vectors from all images in mammography exams.

First, we train a CNN with replaced last layer by a classification layer with only one neuron to solve the binary classification problem (\{cancer, normal\}), as shown in Figure~\ref{fig:single}. The CNN weights are initialized using a pretrained on ImageNet~\cite{russakovsky2015imagenet} model. The trained model serves as a mapping of the image $x_i$ into the feature vector $f(x_i)=z_i$. Second, the CNN block from previous stage are taken to build a \textbf{four}-view \textbf{mam}mograms model classifier based on \textbf{T}ransformer Encoder (MamT$^4$, Figure~\ref{fig:snn}). The CNN extract feature vectors ($z_i^{0}$, $z_i^{1}$, $z_i^2$, $z_i^3$) from both breasts (left and right) and both projections (MLO and CC). During a train in this step weights of CNN block are not trainable. MamT$^4$ predicts labels for the $x_i^{0}$ image, $x_i^1$, $x_i^2$, $x_i^3$ images survey as a additional information (or as a \textit{metadata}).

Each feature vector $z_i$ from EfficientNet-B3 (motivation to utilize EfficientNet-B3 is shown in Section~\ref{sec:experiments}) has length 1536. We divide each vector into 8 \textit{tokens} (the number can be considered as a \textit{hyperparameter}) with a size of 192. Four vectors yield 32 tokens. Learnable position embeddings are added to the token to retain positional information. Similar to BERT’s~\cite{devlin2018bert} and ViT's we add a learnable \texttt{[class]} token, whose state at the output of the TE fed to the MLP head (which is nothing but a linear layer $192\times1$) to get class predictions.

\section{Experiments}
\label{sec:experiments}
We evaluate our proposed MamT$^4$ framework on cancer classification tasks. Experiments are conducted on the \textbf{VinDr-Mammo} dataset~\cite{Nguyen2023-tb}, which was released quite recently and contain 5,000 mammography exams (four images per patient, 20,000 digital mammograms in total). The annotated exams are split into a training set of 4,000 exams and a test set of 1,000 exams. The dataset has BI-RADS classification, the images are classified similarly to the proposed solution~\cite{Shen2019-kp}, specifically: categories 1, 2 -- ``normal'', 4, 5 -- ``cancer'', category 3 is not included. Thus, images have two labels \{cancer, normal\}.

\textbf{Evaluation}.
We use ROC-AUC (the ``gold standard'' for binary classification with neural networks~\cite{Kegeles2020-pd}) and F1 score (the harmonic mean of the Precision and Recall) to measure classification performance. Although we use 5-fold cross-validation to choose the EfficientNet-B3 model, it is not very convenient in the proposed training method which consists of two stages. Due to the following: for the second stage of training MamT$^4$, we would have to keep the split information of the dataset from the first stage of training CNN block. 
Thus, the results in the Table~\ref{tab:crop},~\ref{tab:vindr245} are obtained by training on various 5 seeds.

\subsection{Experimental Setup}
\textbf{Implementation Details}. Our models are trained with the PyTorch framework. In the training process, we set the initial learning rate to $10^{-5}$, and start to attenuate the learning rate when a F1 score on the test set stops improving within 5 epochs. If not specified otherwise, we use FL with the following parameters: $\alpha_0=0.05$, $\alpha_1=0.95$ and $\gamma=2.0$. In the proposed approach, we first randomly apply the cropping method to mammogram. Then the image is resized to $512\times512\times3$. We train the model for 200 epochs with the option to stop early if the F1 score metric does not improve within 10 epochs on the test set. We train the MamT$^4$ model with $L=N=12$ number of TE blocks and heads of MSA. The optimal number of $L$, $N$ could be considered as hyperparameters in future studies.

\begin{table}
  \caption{Encoder performance comparison based on ROC-AUC metrics on VinDr-Mammo dataset.}
  \centering
  \begin{tabular}{@{}lc@{}}
    \toprule
    Encoder            & ROC-AUC  \\
    \midrule
    EfficientNet-B3              & $79.2 \pm 0.8$      \\
    Swin (Tiny)                  & $58.5 \pm 3.2$      \\
    Swin-V2 (Tiny)               & $69.5 \pm 2.3$      \\
    SegFormer-B1                & $61.6 \pm 2.5$      \\
    ResNet-18                    & $78.1 \pm 1.6$      \\
    MobileNet-V3 (Large)        & $76.6 \pm 1.8$      \\
    \bottomrule
  \end{tabular}
  \label{tab:encoders_comparison}
\end{table}

\textbf{Baseline}. 
Our framework is based on the EfficientNet-B3~\cite{DBLP:journals/corr/abs-1905-11946} model, pretrained on ImageNet~\cite{russakovsky2015imagenet}, that is used as an encoder. That model is chosen due to its superior performance on ImageNet, publicly available pre-trained weights, and its optimal balance between high performance and a manageable number of parameters. We performed comparisons with other models on the VinDr-Mammo dataset. Prior to the main experiment, EfficientNet-B3 demonstrates the highest performance, achieving a ROC-AUC score of $79.2\pm0.8$. Models including Swin (Tiny)~\cite{DBLP:journals/corr/abs-2103-14030}, Swin-V2 (Tiny)~\cite{DBLP:journals/corr/abs-2111-09883}, SegFormer-B1~\cite{DBLP:journals/corr/abs-2105-15203}, ResNet-18~\cite{he2016deep}, and MobileNet-V3 (Large)~\cite{howard2019searching} were evaluated. Detailed results for these models, averaged over five independent runs across different data splits, are provided in Table~\ref{tab:encoders_comparison}.

\begin{table}
  \caption{Distribution of datasets split into training and testing subsets, indicating percentage contributions to the cropping dataset.}
  \centering
  \begin{tabular}{@{}lccc@{}}
    \toprule

    & \multicolumn{2}{c}{Split} & \\
    \cmidrule{2-3}
    Dataset & Train (80\%) & Test (20\%) & Total \\
    \midrule
    
    CBIS-DDSM~\cite{Lee2017-qo} & 46 & 16 & 62 (7.8\%) \\
    INbreast~\cite{Moreira2012-lb} & 49 & 14 & 63 (7.9\%) \\
    KAU-BCMD~\cite{Alsolami2021-mu} & 51 & 6 & 57 (7.1\%) \\
    MIAS~\cite{Suckling1994TheMI} & 39 & 13 & 52 (6.5\%) \\
    CMMD~\cite{Cui2021-mr} & 55 & 11 & 66 (8.2\%) \\
    VinDr-Mammo~\cite{Nguyen2023-tb} & 400 & 100 & 500 (62.5\%) \\
    \bottomrule
  \end{tabular}
  \label{tab:crop_datasets}
\end{table}

\textbf{Preprocessing}. We tried different methods for obtaining a breast mask: a classic method based on color selection and a neural network method. The color thresholding method involves two stages. In the first stage, we select a color threshold for the images, setting it at one-quarter of the mean, and then apply the threshold to create a binary mask. After this, we select the largest region by area. This method is computationally effective and simple; however, it has limitations because color values do not usually represent the breast accurately, especially due to the presence of extraneous artifacts such as labels on mammography images. It is also challenging to separate the breast, which is the ROI, from other body parts that may accidentally be present in the image.

The second method is based on neural network predictions using U-Net with a ResNet-34 encoder, pretrained on ImageNet. We use our dataset, where non-professional annotators, under consultation with mammalogists, labeled the borders of the breast without considering projection and laterality. This dataset is compiled using images from six public datasets, including VinDr-Mammo. The proportions of the cropping dataset are shown in Table~\ref{tab:crop_datasets}. We develope a universal model that works for each type of projection. We teste this method and observed a quality improvement compared to the color thresholding method. Our fine-tuned U-Net model shows a performance of 98.6\% mean IoU on the test subset.

After obtaining the mask prediction, we performed a centered crop of the breast and then fed the cropped image into a neural network to predict label.

\textbf{Augmentations}.
We apply augmentations, such as random shuffling, blurring, Gaussian noise, horizontal flips, hue saturation value shifts, sharpening, grid dropouts, grid distortions, coarse dropouts, pixel dropouts. 
Cropping is applied as follows: Each image in the training set is cropped with a probability of 0.5 (with crop aug.) or with a probability of 1 (with crop all), all images in the test set are cropped. 
In the four-view model, we randomly replace $x_i^1$, $x_i^2$, $x_i^3$ with black images (we call it \textit{EmptyImage} augmentation) to indicate model that the dataset may not always contain four images for one patient, with the probability of passing a black image, p, set to 0.2 during a train and set to 0 during a test.

\subsection{Results and Discussion}
\label{sec:results}

\begin{figure}
  \centering
   \includegraphics[width=0.9\linewidth]{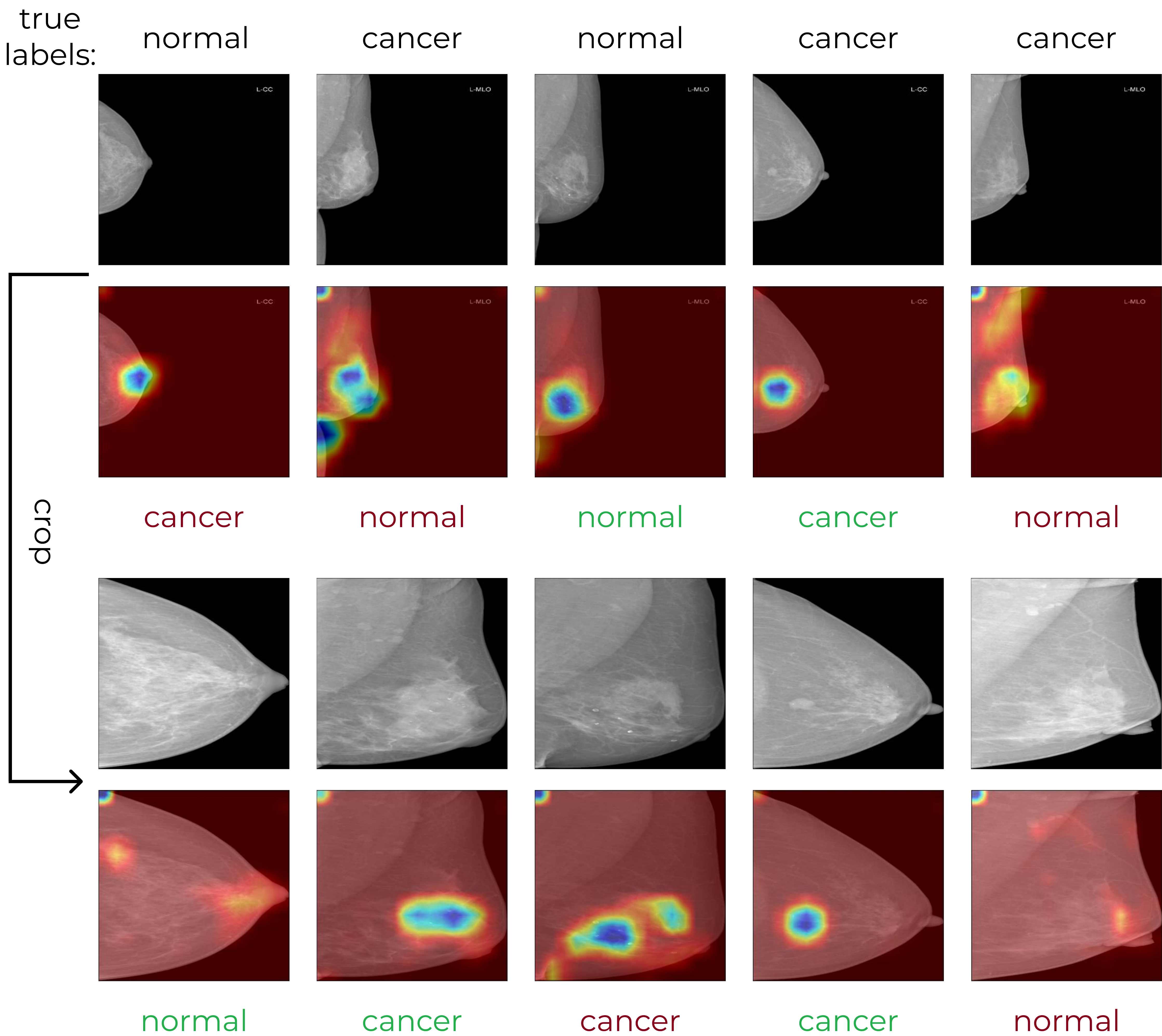}

   \caption{The visualization contrasts the predictions of two models: the model without preprocessing and the model utilizing cropping. In the first two columns, we demonstrate instances where the cropping model made correct predictions on the cropped images, while the model without cropping failed to do so on the original images. The third column presents a less common scenario (34 instances as opposed to 84) where the model without cropping correctly identifies the images as normal, and the cropping model errors. Following that, there is an example where both models correctly identifies cancer. Finally, the last column shows a case where both models make incorrect predictions.}
   \label{fig:gradcam}
\end{figure} 

For testing, we compare a model trained on the original dataset without cropping, which tests ordinary non-cropped images, against a model trained with crop augmentation and evaluated on the same test set but preprocessed with U-Net for cropping. The results, presented in Table~\ref{tab:crop}, indicate that cropping improves the quality of predictions. We also investigate Grad-CAM++~\cite{Chattopadhay_2018} visualizations of the neural networks' predictions (Figure~\ref{fig:gradcam}). Cropping creates a simpler dataset for the deep learning model, containing only relevant information. We also believe that cropping enhances prediction quality because it adjusts the entire breast size to fit within the picture area, thereby achieving scale unification. Furthermore, with the entire image area covered by the breast, the model can better focus on the breast, as the crucial elements, breast tissues, appear larger than in original images.

Table~\ref{tab:crop} shows the performance of MamT$^4$ on the VinDr-Mammo dataset compared with the single-view method (with crop and with crop aug.) and method without using cropping (w/o crop). Comparing the two cropping methods, all three metrics coincide within the standard deviation, so for consistency in the rest of the experiments, ``with crop'' means ``with crop aug.'' The relative improvement of four-view MamT$^4$ (with crop) compared to single-view EfficientNet-B3 (w/o crop) is 4.4\% of ROC-AUC, 11.3\% of F1 score and 5.7\% of F1 score (macro). 
Note that the VinDr-Mammo dataset includes the complete set of four images for each study, with the sole exception of one patient out of 4,000 from the training set, who is manually removed from the dataset (4,000$\to$3,999).
Our approach with EmptyImage augmentation can be used for datasets like the CBIS-DDSM~\cite{Lee2017-qo}, which has no all four mammogram images per patient.

Recently, ROC-AUC of 75.3\% and F1 score (macro) of 76.0\% have been achieved on the VinDr-Mammo dataset~\cite{truong2023delving}. They divide BI-RADS into two classes differently than we do: BI-RADS 2 -- ``normal'', 4, 5 -- ``cancer'', BI-RADS 1 and 3 is not included. Table~\ref{tab:vindr245} show the performance of methods with their method of dividing them into 2 classes. In that case FL has $\alpha_1=0.87$.

We adopt a two-stage approach in our new framework: MamT$^4$, a cancer classification model, is based on the analysis of four views. It is important to note that our framework is versatile and can be adapted to predict different scenarios, such as two projections for a single breast or any required number of images in a specific area. 

\begin{table}
  \caption{Performance comparison of three methods on VinDr-Mammo dataset.}
  \centering
  \begin{tabular}{@{}lccc@{}}
    \toprule
    Method & ROC-AUC & F1 & F1 (macro)  \\
    \midrule
    w/o crop & $79.6 \pm 2.0$ & $44.7 \pm 0.4$ & $71.10 \pm 0.20$  \\
    with crop aug. & $83.8 \pm 0.4$ & $53.2 \pm 1.1$ & $75.5 \pm 0.5$ \\
    with crop all & $82.4 \pm 1.0$ & $52.8 \pm 0.8$ & $75.3 \pm 0.4$ \\
    MamT$^4$ with crop & ${84.0 \pm 1.7}$ & $\mathbf{56.0 \pm 1.3}$ & $\mathbf{76.8 \pm 0.8}$ \\
    \bottomrule
  \end{tabular}
  \label{tab:crop}
\end{table}

\begin{table}
  \caption{Performance comparison of two methods on VinDr-Mammo dataset, when we assume BI-RADS 2 -- ``normal'' and 4, 5 -- ``cancer'', BI-RADS 1,3 is not included.}
  \centering
  \begin{tabular}{@{}lccc@{}}
    \toprule
    Method & ROC-AUC & F1 & F1 (macro)  \\
    \midrule
    with crop & $79.9 \pm 0.9$ & $57.8 \pm 1.1$ & $75.0 \pm 0.7$  \\
    MamT$^4$ with crop & ${80.3 \pm 1.1}$ & $\mathbf{61.0 \pm 1.4}$ & $\mathbf{77.2 \pm 0.7}$  \\
    \bottomrule
  \end{tabular}
  \label{tab:vindr245}
\end{table}

\section{Related Work}

\textbf{Multi-view Analysis}. 
\label{sec:multi_view}
The rationale for using four images simultaneously for prediction is that, in radiologist practices, the symmetry information from other images is utilized to improve the accuracy of decisions. For example, a lesion in one breast rarely appears in the corresponding area in the other breast~\cite{yang2021momminet}. Other multi-view approaches, using two to four images as inputs, have also been proposed~\cite{yang2021momminet,chen2022multi,wang2018breast,carneiro2017automated,li2016survey,nguyen2022novel}. Recent studies indicate that multiple-view approaches improve breast cancer diagnosis~\cite {khan2019multi , geras2017high, truong2023delving}.

\textbf{Applications}.
The proposed method of using a multi-view model to determine the diagnosis of breast cancer from a mammogram can also be applied to other areas of medicine where multiple projections or types of images need to be analyzed for more accurate diagnosis. For example, this technique can be effectively applied to identify the diagnosis of other types of cancer, such as lung, stomach, skin, from different types of scans like computed tomography (CT), magnetic resonance imaging (MRI), and ultrasound, where deep learning techniques are already being actively applied~\cite{lung2, gastric}. 

One example of medical imaging that requires the analysis of multiple projections or types of images for a more accurate diagnosis may be CT in the examination of patients with head injuries. If there is a suspicion of skull fracture or other injuries, doctors may need to review images from different projections to get a complete picture of the injuries and choose the best treatment method. Studies on multi-class semantic segmentation and on detection of abnormalities in traumatic brain injury have already shown their effectiveness and the positive impact of artificial intelligence techniques in optimizing workflow in radiology~\cite{brain_injury, brain2}. Therefore, the proposed method on analyzing multiple projections of CT scans can significantly increase the performance in detecting suspicious areas and thus help clinicians to make a more accurate informed decision on further treatment of the patient.

\section{Conclusion}
\label{sec:conclusion}

Our study achieved metrics on the independent VinDr-Mammo dataset, including a ROC-AUC of 84.0 ± 1.7 and a F1 score of 56.0 ± 1.3. The preprocessing method involved a cropping model that focused on the breast region and removed extraneous artifacts, while also enlarging the breast to the image's size allowing the classification model to better distinguish details. 

Our framework MamT$^4$ utilized multi-view analysis based on Transformer Encoder to improve cancer classification accuracy. Depending on task domain, the number of input images can be increased or decreased. This approach can also be beneficial in various medical imaging applications beyond mammography where different projections of the same object are used, improving accuracy and helping physicians to make informed patient treatment decisions.

\bibliographystyle{IEEEtran}
\bibliography{main}

\end{document}